\documentclass{phb-proc4-auth}
\usepackage{graphicx}
\usepackage{amsmath}
\usepackage{amssymb}
\usepackage{url}
\usepackage{times}

\graphicspath{{.}}

\begin{document}

\begin{frontmatter}

  \journal{SCES '04}
  
  \title{Optical response of electrons in a random potential}

  \author[syd]{Alexander Wei{\ss}e\corauthref{1}},
  \author[hgw]{Gerald Schubert},
  \author[hgw]{Holger Fehske}
  \address[syd]{School of Physics, The University of
    New South Wales, Sydney NSW 2052, Australia} 
  \address[hgw]{Institut f\"ur Physik,
    Ernst-Moritz-Arndt Universit\"at Greifswald, 17487 Greifswald,
    Germany}
  \corauth[1]{Corresponding Author:
    Phone: +61-2-9385-5667, Fax: +61-2-9385-6060,
    Email: aweisse@phys.unsw.edu.au}
  
  \begin{abstract}
    Using our recently developed Chebyshev expansion technique for
    finite-temperature dynamical correlation functions we numerically
    study the AC conductivity $\sigma(\omega)$ of the Anderson model
    on large cubic clusters of up to $100^3$ sites. Extending previous
    results we focus on the role of the boundary conditions and check
    the consistency of the DC limit, $\omega\to 0$, by comparing with
    direct conductance calculations based on a Greens function
    approach in a Landauer B\"uttiker type setup.
  \end{abstract}
  
  
\end{frontmatter}

The numerical calculation of linear response functions is one of the
standard tasks in condensed matter theory and many other areas of
physics. In practice, however, the dimension $N$ of the corresponding
eigenvalue problem usually becomes enormously large. A complete
diagonalisation and a naive evaluation of linear response functions is
prohibitive is such situations, since the required time would scale at
least as $N^3$. In a recent work~\cite{We04} one of us proposed an
advanced Chebyshev expansion method for the calculation of dynamical
correlation functions at finite temperature, which is {\em linear} in
the system size $N$. To yield data for all temperatures and chemical
potentials within a single simulation run marks one of its additional
advantages.  As a particularly interesting application we considered
the optical (AC) conductivity $\sigma(\omega)$ of non-interacting
electrons in a random potential, which can be described by the
well-known Anderson model~\cite{An58}
\begin{equation}\label{defham}
  H = -t \sum_{\langle ij\rangle} 
  \big( c_i^{\dagger} c_j^{} + c_j^{\dagger} c_i^{} \big)
  + \sum_i \epsilon_i^{} c_i^{\dagger} c_i^{}\,.
\end{equation}
Here $c_i^{\dagger}$ ($c_i^{}$) denote fermion creation (annihilation)
operators and $\epsilon_i^{} \in [-W/2, W/2]$ is the uniformly
distributed local potential.

In the present work we extend our recent studies and focus on the
sensitivity of $\sigma(\omega)$ to changes in the boundary conditions.
In addition, we check the consistency of the $\omega\to 0$ limit by
comparing extrapolations of the AC data and direct calculations of
the DC conductance combined with finite size scaling.  

Within linear response theory the real part of the optical
conductivity is given by
\begin{equation}\label{defsigma}
  \sigma(\omega) = \sum_{n,m}
  \frac{|\langle n|J_x| m\rangle|^2}{\omega L^d} [f(E_m) - f(E_n)]\,
  \delta(\omega - \omega_{nm})\,,
\end{equation}
where $|n\rangle$ and $|m\rangle$ denote eigenstates of the
Hamiltonian with energies $E_n$ and $E_m$, $\omega_{nm}=E_n-E_m$,
$f(E) = [\exp(\beta(E-\mu)) + 1]^{-1}$ is the Fermi function, and $J_x
= -\text{i} t \sum_{i} (c_i^{\dagger} c_{i+x}^{} - c_{i+x}^{\dagger}
c_i^{})$ the $x$-component of the current operator. Even at zero
temperature Eq.~\eqref{defsigma} involves a summation over matrix
elements between {\em all} one-particle eigenstates of $H$, which for
reasonably large systems can hardly be calculated explicitly.
Consequently, until now, the number of numerical attempts to this
problem is very limited~\cite{AG78*SM85*HGF87*LS94b*SN99}.
Rewriting $\sigma(\omega)$ in terms
of Fermi integrals over a matrix element density $j(x,y)$, which
itself is independent of temperature and chemical potential,
our new approach~\cite{We04} is based on the observation that the
latter can be efficiently calculated with a {\em two-dimensional}
generalisation of Chebyshev expansion and the Kernel Polynomial
Method~\cite{SR94}. This leads to an $O(N)$ algorithm and allows the
study of systems with as many as $100^3$ sites.

\begin{figure}
  \begin{center}
    \includegraphics[width=0.9\linewidth]{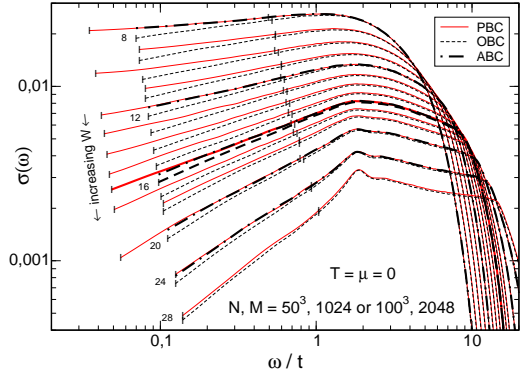}
  \end{center}
  \caption{Optical conductivity of the 3D Anderson 
    model at $T=0$ and $\mu=0$ (band centre) for increasing disorder
    $W$. Bold lines mark $W/t = 16$, which approximately
    corresponds to the critical disorder. Most curves correspond to
    $N=50^3$ sites, expansion order $M=1024$ and averaging over $240$  
    samples; sets that extend to lower $\omega$ are based on $N=100^3$, 
    $M=2048$ and $400$ samples.}\label{figsigma}
\end{figure}

\begin{figure}
  \begin{center}
    \includegraphics[width=0.9\linewidth]{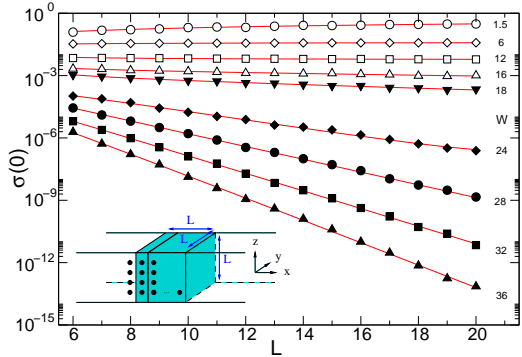}
  \end{center}
  \caption{Finite size scaling of the DC conductivity calculated 
    directly~\protect\cite{Ve99} within the Landauer B\"uttiker type geometry
    shown in the inset.}\label{figsnull}
\end{figure}

\begin{figure}
  \begin{center}
    \includegraphics[width=\linewidth]{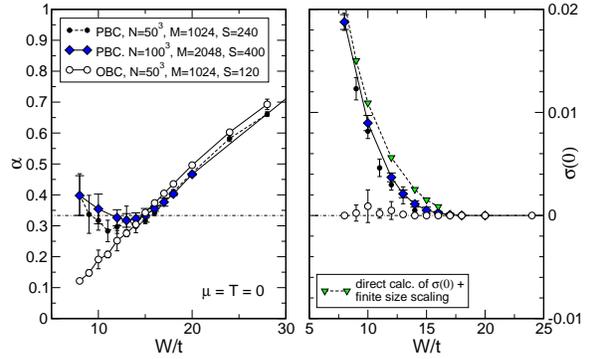}
  \end{center}
  \caption{Exponent $\alpha$ and DC conductivity 
    $\sigma(0)$ obtained from fits of the low-frequency conductivity
    to $\sigma(\omega) = \sigma(0) + C\,\omega^{\alpha}$ (vertical
    bars in Fig.~\ref{figsigma} mark the underlying frequency range).
    Error bars are estimated by slightly varying $\mu\in [-0.05W,0.05W]$.
  }\label{figfits}
\end{figure}

The good quality of the obtained conductivity data allows for tests of
the various analytical predictions for the finite frequency behaviour.
In particular, for $d>2$ dimensions we expect $\Delta\sigma =
\sigma(\omega) - \sigma(0) \sim \omega^{(d-2)/2}$ in the metallic
phase, $\sigma(\omega) \sim \omega^{(d-2)/d}$ at the metal-insulator
transition, and $\sigma(\omega)\sim \omega^2$ far inside the
insulating phase~\cite{Mo67*We76*OW79*SA81*Sh82}.
Figures~\ref{figsigma} and~\ref{figfits} illustrate that for low
frequency $\sigma(\omega)$ is indeed well described by a power law
$\sigma(0) + C\,\omega^{\alpha}$. As expected, for samples with open
boundary conditions (OBC) the extrapolated DC contribution $\sigma(0)$
is always zero, whereas for periodic (PBC) and anti-periodic (ABC)
boundary conditions we observe a continuous metal-insulator transition
at $W/t\approx 16$. As a further check we also calculated $\sigma(0)$
directly using a numerical Greens function approach within a Landauer
B\"uttiker type geometry~\cite{Ve99} and the finite size scaling
ansatz $\sigma(0)=L e^{-L/\lambda}/(a+L)$ (cf. Fig.~\ref{figsnull}).
Interestingly, within both approaches the resulting critical exponent
for $\sigma(0)$ seems to be of the order of 2, larger than the
expected $\nu\approx 1.57$, a puzzle which certainly requires further
examination. Focussing on the exponent $\alpha$, our PBC and ABC data
appears to be consistent with the analytic predictions, i.e.,
$\alpha\to 0.5$ for weak disorder, $\alpha\approx 1/3$ at the critical
disorder, and eventually $\alpha\to 2$ for very strong disorder. Note,
however, that for OBC $\alpha$ is a simple, monotonously increasing
function of $W$.  The above results show, that the new technique is a
reliable numerical tool for large scale calculations of dynamical
correlations functions.

We acknowledge financial support by the Australian Research Council,
calculations were performed at APAC Canberra, ac3 Sydney and NIC
J\"ulich.


\end{document}